\documentclass[showpacs,preprint]{revtex4}
\usepackage{amsmath}
\usepackage{graphicx}
\usepackage{amsfonts, amssymb}
\usepackage{gensymb}
\usepackage{cool}
\usepackage{datetime}

\def\be{\begin{equation}}
\def\ee{\end{equation}}

\def\Hef{\vec{H}_{\mbox{{\tiny eff}}}}

\def\hef{\vec{h}_{\mbox{{\tiny eff}}}}

\def\kpar{K_{\parallel}}
\def\kperp{K_{\bot}}
\begin{document}

\title{ Speed of  field driven domain walls  in   nanowires with large  transverse magnetic anisotropy }

\author{M. C. Depassier}
\email{mcdepass@uc.cl}
\affiliation{Instituto de F\'\i sica, Pontificia Universidad Cat\'olica de Chile \\ Casilla 306, Santiago 22, Chile}

\date{ \today}

\begin{abstract}
Recent  analytical and numerical work on  field driven domain wall propagation  in  nanowires  has shown that  for large transverse anisotropy and sufficiently large applied fields the Walker profile becomes unstable before the breakdown field, giving way to a slower stationary domain wall.
 We perform an asymptotic expansion of  the Landau Lifshitz Gilbert equation for  large transverse magnetic anisotropy and show that the asymptotic dynamics reproduces this behavior. At low applied field the speed increases linearly with the field and the profile is the classic Landau profile.  Beyond a critical value of the applied field the domain wall slows down.  The appearance of a slower domain wall profile in the asymptotic dynamics is due to a transition from a pushed to a pulled front of a reaction diffusion equation.   \end{abstract}


\pacs{75.78.-n, 75.78.Fg }

\maketitle

Magnetic domain wall propagation is an active area of research both as an interesting physical phenomenon as well as for its possible applications in logic devices, magnetic memory elements and others \cite{Stamps2014}.      The dynamics of magnetic domain walls is described by the Landau Lifshitz Gilbert (LLG) equation \cite{Landau1935,Gilbert1956} which cannot be solved analytically except in very special cases.    For an infinite medium with uniaxial anisotropy and an external field applied along the symmetry axis, the Walker solution \cite{ScWa1974}   provides the best known analytical expression for the  profile and speed of the domain wall.  The exact Walker solution, predicts that the speed increases linearly with the field up to a critical field $H_w$.  Above this value  a sudden drop in velocity and  an irregular precessing  motion of the magnetization appears.  Field induced domain wall propagation in thin films and nanowires  has been examined with greater detail in recent work.  The numerical study \cite{WiVe2010} showed that  depending on the  relative magnitude of hard axis anisotropy  different scenarios arise.   For small hard axis anisotropy the Walker solution is realized.  For sufficiently large values of the hard axis anisotropy the Walker breakdown does not occur.  There is a slowdown  of the domain wall due to spin wave emission and no sudden drop in speed. For the largest values considered in \cite{WiVe2010} the domain wall speed changes from a regime of linear growth with the applied   field to a regime of slower growth with increasing applied field.   This last behavior is observed when both the exchange constant and uniaxial anisotropy are much smaller than hard axis anisotropy. Further numerical studies \cite{WaYa2012,WaWa2014} analyze in detail the nature of the spin waves emitted and distinguish two scenarios, depending on the relative values of the exchange  and anisotropy constants. The parameter ranges studied in \cite{WiVe2010} and \cite{WaWa2014} differ, however, in both cases the Walker breakdown is not observed when the transverse anisotropy is sufficiently large. The stability of the Walker solution with respect to small perturbations  has been studied  recently \cite{HuWa2013} using dynamical systems techniques. The analysis of the spectrum of a perturbation to the Walker solution shows that it may become absolutely or convectively  unstable before the breakdown field. This instability is found  for sufficiently large transverse anisotropy and for fields larger than a critical value.

The purpose of this work is to study the dynamics of the LLG equation  for a nanowire when the transverse anisotropy is large by means of an asymptotic expansion. The asymptotic expansion captures the slower relaxation dynamics of the domain wall and filters out the fast spin waves \cite{GaCeE2001}.  We find that the leading order asymptotic dynamics predicts a transition from a Walker type regime to a regime with slower domain wall motion. In leading order the dynamics of the in plane magnetization obeys a reaction diffusion equation, and the perpendicular magnetization is slaved to the in-plane components. The slowdown of the domain wall in this asymptotic regime appears as a transition from a pushed to a pulled front at a critical value of the applied  field. 
 
The starting point of the calculation  is the LLG  equation for the magnetization. The material has magnetization $\vec M= M_s \vec m$ where $M_s$ is the saturation magnetization and $\vec m = (m_1,m_2,m_3) $ is a unit vector along the direction of magnetization. 
The dynamic evolution of the magnetization  is governed  by the LLG equation, 
\be\label{LLG}
 \frac{d \vec M }{ d t} = - \gamma_0  \vec M \times \Hef + \alpha\frac{\vec M}{M_s} \times \frac{d \vec M }{ d t}
 \ee

where $\Hef$ is the effective magnetic field,  $\gamma_0=  | \gamma | \mu_0$,   $\gamma$ is the gyromagnetic ratio of the electron and  $\mu_0$ is the magnetic permeability of 
vacuum. The constant  $\alpha>0$ is the dimensionless phenomenological Gilbert damping coefficient. 
We consider a thin and narrow  film in the $(x,y)$ plane, with the easy axis along its length.   The strip is subject to an applied magnetic field along the easy axis $\vec H_a = H_a \hat x$.  The film is thin and narrow so that the magnetization may be assumed \cite{PoDo2004} to depend on the easy axis coordinate, $\vec M(x,y,z) = \vec M(x)$. In addition, in this geometry  the demagnetizing field  has a local representation as an effective perpendicular anisotropy so that, as in \cite{WiVe2010,WaYa2012,HuWa2013, WaWa2014}, the effective magnetic field  is given by
\be \label{H1}
\Hef = H_a \hat x + \frac{C_{\rm ex}}{\mu_0 M_s^2}  \frac{\partial^2 \vec M}{\partial x^2} + \frac{2 K_u}{\mu_0 M_s^2} M_1\hat x -  \frac{2 K_d}{\mu_0 M_s^2} M_3 \hat z,
\ee
where $C_{ex}$ is the exchange constant, $K_u$ the easy axis uniaxial anisotropy and $K_d$ the perpendicular anisotropy.

Introducing $M_s$ as unit of magnetic field, and introducing the dimensionless space  and time variables   $\xi =  x \sqrt{K_u/C_{ex}}$  and $\tau= \mu_0 |\gamma | M_s t$  we rewrite equations (\ref{LLG}) and (\ref{H1}) in dimensionless form

\be\label{LLG1}
 \frac{d \vec m }{ d \tau} = -   \vec m \times \hef + \alpha \vec m  \times \frac{d \vec m }{ d\tau }
 \ee
 with
\be \label{hef}
\hef= h_a \hat x + \frac{1}{2} \kpar  \frac{\partial^2 \vec m}{\partial \xi^2} + \kpar m_1 \hat x - \kperp m_3 \hat z.
\ee
where $h_a$ is the dimensionless applied field and the dimensionless numbers that have appeared are $\kpar= 2 K_u/(\mu_0 M_s^2)$, $\kperp= 2 K_d/(\mu_0 M_s^2)$. 
Equations (\ref{LLG1})  and (\ref{hef})  describe the dynamics of the problem.

We are interested in the case of a perpendicular anisotropy  much larger than the uniaxial in plane anisotropy.   In this situation   the perpendicular magnetization $m_3$ will be smaller than the in plane components.  We will also assume that the dimensionless applied field is weak.  
We search then for a solution of the LLG equation  in the asymptotic limit   $h_a << \kperp$, $\kpar << \kperp$  and therefore,  $m_1, m_2 >>  m_3$.
Let then
$$
m_1 = m_{10} + \epsilon m_{11} +\ldots, \qquad  m_2 = m_{20} + \epsilon m_{21} +\ldots,  \qquad m_3 = \epsilon  m_{30} + \epsilon^2 m_{31} +\ldots,
$$
where $\epsilon$ is a small quantity. Since the perpendicular anisotropy is larger than the uniaxial anisotropy and the applied field is weak, we introduce the scaling  $\kpar = \epsilon \tilde \kpar$, $h_a= \epsilon \tilde h_a$  with $\kperp$ of order one. The  components of the effective magnetic field,  $\hef = (h_1,h_2,h_3)$   become then
$
h_i = \epsilon h_{i0} + \epsilon^2 h_{i1} + \ldots$ 
with the leading order components  given by
\be\label{h0}
h_{10} =  \tilde h_a + \frac{1}{2} \tilde \kpar \frac{\partial^2  m_{10}}{\partial \xi^2} +  \kpar m_{10}, \qquad h_{20} =  \frac{1}{2} \tilde \kpar \frac{\partial^2  m_{20}}{\partial \xi^2} 
\quad {\rm and} \quad h_{30} = - \kperp m_{30}.
\ee
 Furthermore we introduce a slow time scale $s = \epsilon \tau$ and notice that the leading order components of the in plane magnetization satisfy
 \be \label{norm}
m_{10}^2 + m_{20}^2 = 1  - O(\epsilon^2).\ee
Introducing these scalings in Eq.(\ref{LLG1}) and expanding in $\epsilon$  one obtains
\begin{subequations}
\begin{align}
\label{una}
\frac{ \partial  m_{10}}{\partial s}  &= - m_{20} h_{30} + O(\epsilon), \\
\label{dos}
\frac{ \partial  m_{20}}{\partial s} &= m_{10}h_{30}  + O(\epsilon), \\
\label{tres}
0 &= -m_{10} h_{20} + m_{20} h_{10} +\alpha  ( m_{10}  \frac{ \partial  m_{20}}{\partial s}  - m_{20}    \frac{ \partial  m_{10}}{\partial s}  ) + O(\epsilon).
\end{align}
\end{subequations}
Substituting (\ref{una}) and (\ref{dos}) into (\ref{tres}) and using (\ref{norm}) we find that in leading order, 
\be\label{h3}
h_{30} = \frac{1}{\alpha} (m_{10}h_{20} - m_{20} h_{10})
\ee
and equations (\ref{una})  and (\ref{dos}) become
\begin{subequations} \label{m0}
\begin{align} 
\frac{ \partial  m_{10}}{\partial s} &= -  \frac{m_{20}}{\alpha}  (m_{10}h_{20} - m_{20} h_{10}) \label{m10} \\
\frac{ \partial  m_{20}}{\partial s} &=  \frac{m_{10}}{\alpha}  (m_{10}h_{20} - m_{20} h_{10}).  \label{m20}
\end{align}
\end{subequations}
Because of (\ref{norm}) we  can write $m_{10} = \cos\theta$, $m_{20}=\sin\theta$. Using (\ref{h0}) in equations  (\ref{h3}-\ref{m0})  we obtain 
\begin{align}\label{casi}
\alpha     \frac{ \partial  \theta}{\partial s} &=   \frac{1}{2} \tilde \kpar  \theta_{\xi\xi} - \sin\theta (\tilde h_a  + \tilde \kpar \cos\theta ) \\
 m_{30} &=  - \frac{1}{\kperp}   \frac{ \partial  \theta}{\partial s}.
 \end{align}
Finally going back to the  unscaled time variable $\tau$  and parameters $\kpar, \kperp$, we write the leading order magnetization components as 
\begin{subequations} \label{final}
\begin{align}
&  m_1= \cos\theta, \qquad  m_2 = \sin\theta, \qquad  m_3 = - \frac{1}{\kperp} \frac{\partial\theta}{\partial \tau} \quad \textrm{where}\\
\label{rd1} & \alpha \frac{\partial \theta}{\partial \tau} = \frac{\kpar}{2}  \theta_{\xi\xi} - \sin\theta ( h_a  +  \kpar \cos\theta). \end{align}\end{subequations}

Equations (\ref{final}) show that the leading order dynamics is determined by the equation for the in-plane magnetization components, the perpendicular magnetization is slaved to the tangential magnetization.
 Equation (\ref{rd1}) is the well studied reaction diffusion equation, for which we know that an initial perturbation to an unstable state evolves into the monotonic  front of minimal speed \cite{AW78}. In order to render (\ref{rd1})  into the standard form we introduce the dependent variable $u$ defined by $\theta = \pi (1-u)$ which satisfies
\be \label{rd2}
\alpha u_\tau= D  u_{\xi\xi} + f(u), \qquad \textrm{with} \quad f(u) = \frac{1}{\pi} \sin\pi u (h_a - \kpar \cos \pi u).
\ee
The diffusion constant $D= \kpar/2$ and the reaction term $f$ satisifies $f(u)>0$ in (0,1), $f(0) = f(1) =0$. A small perturbation to the unstable state $u=0$ ($\theta=\pi$) evolves into  a traveling monotonic front of minimal speed $c^*$ \cite{KPP37,AW78} that joins the unstable state to the stable state $u=1   (\theta=0)$.  The minimal speed can be obtained from a variational principle \cite{BD96} and is bounded by\cite{AW78}
\be\label{bounds}
c_{\text{KPP}} \equiv \frac{2}{\alpha}  \sqrt{D f'(0)} < c^* < \frac{2}{\alpha}  \sqrt{ D \sup f(u)/u }.
 \ee
When the upper and lower bounds coincide the speed is exactly $c_{\text{KPP}} $ and  the traveling front is called a KPP or pulled front. 

In the present problem Eq. (\ref{rd2}) has the exact traveling front solution
\be
\label{eq:pushed}
u(\xi,\tau)= \frac{2}{\pi} \arctan\left[\text{e}^{- \frac{\kpar}{D}( \xi - c_N  \tau)} \right], \qquad \text{where}\quad  c_{\text N} = \frac{h_a}{\alpha}  \sqrt{\frac{D}{\kpar}}.
\ee
This solution is not a KPP front, it is a so called pushed front. This is the front into which an initial condition will evolve it is effectively  the front of minimal speed. It is not difficult to verify that as $h_a$ increases this is not the speed of the front.   For $h_a  \ge 4 \kpar$ the upper and lower bounds in (\ref{bounds}) coincide and the speed of the front must be  the KPP value. The transition from a pushed to a pulled front may occur before the upper and lower bounds coincide. In this problem for which there is an exact solution we know that the transition will occur when $c_{\text N} = c_{\text{KPP}}.  $ That is, 
$$
c=  \left \{\begin{array}{ll}   \frac{h_a}{\alpha}  \sqrt{\frac{D}{\kpar}} & {\rm if} \,  h_a \le 2 \kpar \\
 \frac{2}{\alpha}  \sqrt{ D  (h_a - \kpar)}  & {\rm if} \,  h_a > 2 \kpar \end{array} \right.
$$

Going back to the physical  variables, we have then that the speed of the domain wall is given by 
\begin{equation}\label{speed}
v= \left\{
  \begin{array}{lr}
 \frac{1}{\alpha} \sqrt{\frac{C_{\text ex}}{ 2 K_u}}  \mu_0  |\gamma | H_a   \qquad   &\text{if} \, \, H_a <  \frac{ 4 K_u}{\mu_0 M_s}   \\
 \frac{2 |\gamma | \sqrt{C_{\text{ex}}}}{\alpha M_s}   \sqrt{ \mu_0 M_s H_a - 2 K_u}  \qquad  &\text{if} \,\,  H_a >  \frac{ 4 K_u}{\mu_0 M_s}  
\end{array}\right.
\end{equation}

In the small field regime  $H_a <  H_c=  4 K_u/ (\mu_0 M_s)$ the magnetization profile is obtained from (\ref{eq:pushed}) and it is given by
\be\label{profile}
m_1 = \tanh \left[ - \frac{(x - v t) }{\Delta}\right], \quad m_2 = \sech  \left[  \frac{(x - v t)}{\Delta} \right] , \quad m_3 = \frac{\mu_0 M_s H_a}{\sqrt{2} \alpha K_d}  \sech \left[  \frac{ (x - v t)}{\Delta} \right], 
\ee
where the domain wall width is given by $\Delta = (1/2) \sqrt{C_{ex}/K_u}$. For an applied field larger than $H_c$ we cannot construct an explicit solution for the magnetization, we can only determine the speed. The general theory of reaction diffusion equations guarantees that it is a monotonic decaying front  similar  in shape to (\ref{eq:pushed}). 
Equations (\ref{speed}) and (\ref{profile}) constitute  our main result.

These results,  obtained from the LLG equation  in the case  $\kpar << \kperp$ and for a weak applied field, explain qualitatively the results of the numerical simulations \cite{WiVe2010,WaYa2012,WaWa2014} and of the stability results \cite{HuWa2013}. At low fields the speed of the front is proportional to the applied field $H_a$ and inversely proportional to the damping coefficient $\alpha$. The magnetization profile and the speed share the main features of the  Walker solution, the velocity shows  linear dependence on the applied field and inverse proportionality  on the damping constant $\alpha$.   The Walker breakdown field $H_W= \alpha \kperp/2$ is of order one, and therefore large compared to the transition field $H_c$.  Thus, we recover the behavior described in \cite{WiVe2010}-\cite{HuWa2013}:   for sufficiently large perpendicular anisotropy the Walker solution loses stability before the breakdown field to a slower moving domain wall. When the applied field is weak and $\kpar<<\kperp$ the numerical integrations in \cite{WiVe2010,WaWa2014} show that the speed increases slowly with the field once the Walker solution loses stability,  in agreement with the results found in this work. 

The asymptotic approach that we have used is based on \cite{GaCeE2001}, where the numerical simulations, (although for a different  demagnetizing field), show that the asymptotic dynamics reproduces the relaxation dynamics of the full LLG equation, filtering out the spin waves. Reaction diffusion dynamics has also been encountered in thin nanotubes \cite{GoRoSl2014}, where the Walker breakdown is not observed.  A transition from a fast to a slower  domain wall also occurs in thin nanotubes as reported in \cite{YaKa2011,De2014}.  The asymptotic dynamics of the LLG equation has been studied by several authors in different limiting parameter ranges, and wave type motion governed by other evolution equations has been derived \cite{Mik1981,CaMeOt2007}. In the present problem we have chosen a parameter regime for which recent numerical work has been performed and found qualitative agreement with the results reported in them.

 \begin{acknowledgments}
 This work has been partially supported by Fondecyt (Chile) projects   1120836 and 
 Iniciativa Cient\'\i fica Milenio, ICM (Chile), through the Millenium Nucleus RC120002.
\end{acknowledgments}

\end{document}